\documentclass[twocolumn,showpacs,amsmath,amssymb]{revtex4}
\usepackage{graphicx}
\begin{document}
\DeclareGraphicsExtensions{.eps, .jpg}
\bibliographystyle{prsty}
\input epsf

\title {Sub-THz electrodynamics of  the graphene-like superconductor  CaAlSi}
\author{S. Lupi$^{1}$, L. Baldassarre$^{1}$, M. Ortolani$^{2}$, C. Mirri$^{1}$,  U. Schade$^{2}$, R. Sopracase$^{1}$, T. Tamegai$^{3}$,  R. Fittipaldi,$^{4}$, A. Vecchione$^{4}$, and P. Calvani$^{1}$}
\affiliation{$^{1}$CNR-INFM "Coherentia" and Dipartimento di Fisica, Universit\`a di Roma La Sapienza, Piazzale A. Moro, 2, 00185 Roma, Italy}\
\affiliation{$^{2}$Berliner Elektronenspeicherring-Gesellshaft f\"ur 
Synchrotronstrahlung m.b.H., Albert-Einstein Strasse 15, D-12489 Berlin, 
Germany}\
\affiliation{$^{3}$ Department of Applied Physics, The University of Tokyo, 7-3-1 Hongo, Bunkyo-ku, Tokyo 113-8656, Japan}
\affiliation{$^{4}$ CNR-INFM Laboratorio Regionale Supermat and Dipartimento di Fisica "E. R. Caianiello", Via S. Allende, 84081 Baronissi Salerno, Italy}\

\date{\today}

\begin{abstract}
We report the first optical study of CaAlSi, a superconductor which displays both the crystal structure of MgB$_2$ and the electronic band structure of  intercalated graphites. The reflectivity of a CaAlSi single crystal was measured down to sub-THz frequencies and to 3.3 K, with the use of Coherent Synchrotron Radiation. A  superconducting gap  in the hexagonal planes, two gaps  along the $c$ axis were found and measured, as expected from the structure of the CaAlSi Fermi surface. The anisotropic optical parameters of the normal state were also determined.
\end{abstract}

\maketitle

In the last few years, much interest has been devoted to the superconducting (SC) properties of graphene-like compounds like  MgB$_2$, up to now  the BCS superconductor with the highest $T_c$, and the intercalated graphites CaC$_6$ and YbC$_6$. All these compounds have  $ab$ planes with hexagonal symmetry, although differently stacked along the $c$ axis. This  reflects into a $k$ space with electronic $\sigma$, $\pi$, and interlayer ($IL$) bands which are differently filled in different members of the family\cite{Boeri07}.  MgB$_2$ is hole-doped, with the Fermi level crossing a two-dimensional (2D) $\sigma$ band and a 3D $\pi$ band \cite{Nagamatsu}.  As a result, it presents two distinct SC gaps. In turn CaAlSi\cite{Imai}  is electron-doped like  CaC$_6$: the $\sigma$ bands are  completely filled,  and the bottom of the $IL$ band crosses the Fermi level together with the $\pi$-bands\cite{Mazin,Boeri,Huang}, again as  in intercalated graphites.The order parameter of CaAlSi is  reported to have a $s$-anisotropic symmetry\cite{Prozorov}, but it is not clear if there are one or two gaps. The available experimental data are not conclusive under this respect. While angle resolved photoemission (ARPES)\cite{Tsuda04} distinguished in CaAlSi one isotropic gap of about 1.2 meV = 4.2 k$_B$T$_c$, muon spin rotation ($\mu$SR) data\cite{Kuroiwa} suggested a highly anisotropic gap, or possibly two distinct gaps like in MgB$_2$. Penetration depth measurements\cite{Prozorov}  are interpreted in terms of a single, moderately anisotropic $s$-wave gap. 
Whether the gap is single or not is an important issue, as  two  coexisting order parameters may lead  to an enhancement of $T_c$  \cite{Suhl,Tsuda05}. Recently, $\mu$SR experiments have reported the existence of a second gap  in the $ab$ planes of high $T_c$ superconductors \cite{Khasanov}.

\begin{figure}[tbp]
    \epsfxsize=8cm \epsfbox {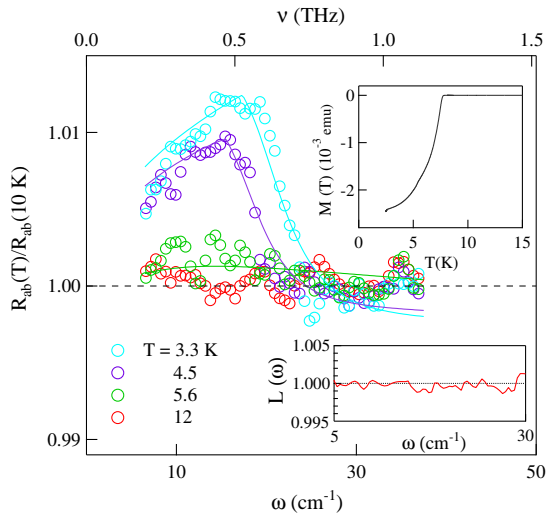}
     \caption{Color online. Ratio between the sub-THz reflectivity of the $ab$ planes below $T_c$ and in the normal phase at 10 K (circles). The lines are fits based on a BCS reflectivity below $T_c$ and on a Hagen-Rubens model at 10 K. The upper inset shows the magnetic moment measured by cooling the sample in a field of 10 Oe, with a superconducting transition at $T_c$ = 6.7 K. In the lower inset, $L(\omega)$ is the ratio between two subsequent spectra collected by CSR, with 256 scans each and a resolution of 0.5 cm$^{-1}$. The error in the sub-THz range  is $\sim 10^{-3}$.}
\label{ratio-ab}
\end{figure}

Infrared spectroscopy (IRS) allows the direct measurement of the optical gap of a superconductor along different crystal axes, with an energy resolution higher than in ARPES. This can be done also in low-$T_c$ materials\cite{Ortolani} with the needed signal-to-noise ratio, with  a conventional  interferometer and a liquid-He bolometer, provided that  the source is Coherent Synchrotron Radiation (CSR)\cite{Abo-Bakr}. Here we apply this technique to a study of the electrodynamics of  CaAlSi. As shown above, this compound may be considered as a model system for the graphene-like superconductors. Moreover, unlike for other members of the  family, CaAlSi crystals can be  grown with all the three dimensions  large enough for optical measurements in the millimeter-wavelength domain.
The  sample  was  a single crystal of 2x4.5x3 mm$^3$. It was grown by the floating-zone method and characterized as described in Ref. \onlinecite{Tamegai03}.  X-ray diffraction  showed an admixture of the  5-fold and 6-fold superstructures reported\cite{Sagayama} for CaAlSi. The cell parameter $c$  was 0.4388 nm. Both surfaces were accurately polished with a  diamond powder having a grain size of 0.5 $\mu$m. After polishing, their alignment with respect to the crystal axes was controlled once again by Electron BackScatter Diffraction (EBSD) in different points. The sample magnetic moment $M(T)$ was also measured after the polishing, and reported in the inset of Fig. \ref{ratio-ab}. It shows the SC transition with an onset at 6.7 K. This temperature, where the resistivity is already equal to zero, is the best estimate of $T_c$ as $M(T)$ is a bulk quantity like the infrared conductivity. 

By assuming that CaAlSi is a BCS superconductor, we may expect a superconducting gap $2\Delta \simeq 3.5 k_B T_c/hc \simeq$ 16 cm$^{-1} \simeq$ 0.5 THz. In order to measure the reflectivity within a few parts per thousand at sub-THz frequencies, we illuminated the sample with the CSR extracted from the electron storage ring BESSY, working in the so-called low-$\alpha$ mode \cite{Abo-Bakr} at a beam current $i \sim 20$ mA. CSR is free of thermal noise and, in the 
sub-THz range, is more brilliant than any other broad-band radiation source by 
at least three orders of magnitude \cite{Abo-Bakr}. The reproducibility between two subsequent spectra is within $\sim$ 0.1\% below 1 THz, as shown in the lower inset of Fig.\ \ref{ratio-ab}. In that range we could thus measure at different temperatures the radiation intensity reflected by the sample $I_R(T)$ as described in Ref. \onlinecite{Ortolani}, in such a way that any effect of diffraction, change in the ring current, and mechanical instability, could be ruled out. The very high reflectivity of the sample and its thickness also excluded any possibility of multiple reflections, as confirmed by the absence of fringes in all frequency domains. By using a commercial interferometer and a bolometer working at $1.6$ K,  we then determined the ratio $I_R(T)/I_R(10 K)$ with an error $\sim$ 0.2 \%. In the present experimental conditions, it coincides with the reflectivity ratio  $R(T)/R(10 K)$ between the reflectivity of CaAlSi in the SC and  in the normal phase. This function is expected to exhibit a peak\cite{Basov} at $\omega = 2\Delta$. 

\begin{figure}[tbp]
    \epsfxsize=8cm \epsfbox {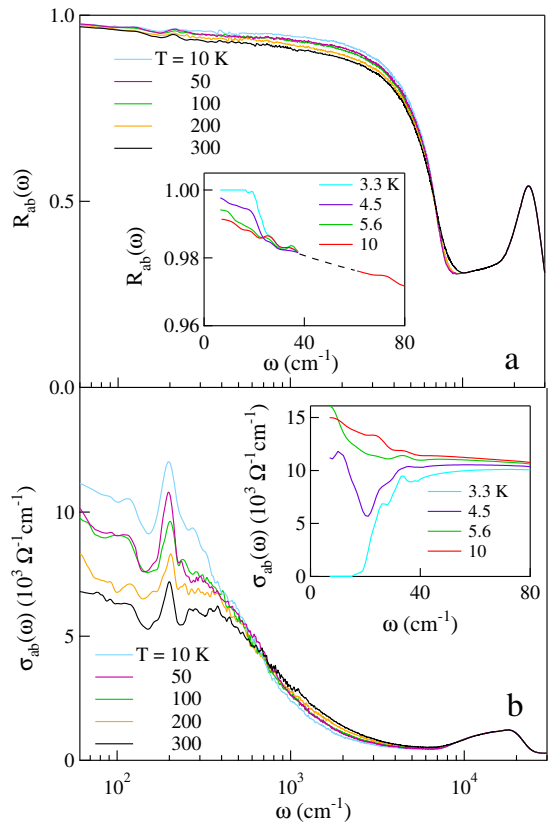}
     \caption{Color online. Optical  response of the hexagonal planes of CaAlSi  a) absolute reflectivity  at different temperatures; in the inset, reflectivity in the THz range above and below $T_c$, as reconstructed from the ratios in Fig.\ \ref{ratio-ab}; b) optical conductivity from the reflectivity in a; in the inset, the conductivity of the hexagonal planes in the sub-THz range.}
\label{plane}
\end{figure}

The ratio $R(T)/R(10 K)$ is reported in Fig.\ \ref{ratio-ab} for the radiation polarized in the hexagonal sheets at $T$, both below and above $T_c$. As $R(10 K) \simeq$ 0.99 at low $\omega$ (see Fig.\ \ref{plane}-a), the total variation of $R_{ab}(T)/R_{ab}(10 K)$ between $T$ = 3.3 and $T$ = 8 K is just slightly more than 1\%. In Fig. \ref{ratio-ab} the curves at $T < T_c$ exhibit the expected peak at $2\Delta_{ab}(T)$, while for $T > T_c$ (12 K), the above ratio is equal to 1 at any $\omega$ within the noise. At 3.3 K the peak frequency is about 17 cm$^{-1}$, which gives $2hc\Delta_{ab}/k_BT_c \simeq 3.8$. It is then reasonable to use the BCS theory to fit the data. We modeled the optical conductivity $\sigma_{ab}(\omega)$, of the hexagonal planes in the normal state, by a conventional Drude model with the plasma frequency $\omega^0_{ab}$ and the scattering rate $\Gamma_{ab}$ determined by a fit to the normal state (T= 10 K) optical conductivity (see below). In the SC state we used the Mattis-Bardeen model with a fixed  $T_c = 6.7$ K and $\Delta_{ab}$ as a free parameter. The curves $R_{ab}(T)/R_{ab}$(10 K) calculated in this way are also reported in Fig. \ref{ratio-ab}. The fit is good at the three temperatures and provides $2\Delta_{ab}$ = 15 and 17.5 cm$^{-1}$ at 4.5 and 3.3 K, respectively. This leads to an extrapolated value\cite{DresselGruner} $2\Delta_{ab} (0)$ = 19.0 $\pm$ 1.5 cm$^{-1}$, or $2hc\Delta_{ab}(0)/k_BT_c = 4.1\pm 0.4$, a value which confirms - with the higher resolution provided by infrared spectroscopy - a previous determination by ARPES \cite{Tsuda04}. This result suggests that CaAlSi is a BCS superconductor with moderately strong electron-phonon coupling. On the basis of our fits, a single gap well describes the superconducting transition in the hexagonal planes.

Once determined the above reflectivity ratios in the sub-THz region, we measured the absolute reflectivity $R_{ab}(T)$ at any $T$ from 60 to 12000 cm$^{-1}$, where $R$ depends on $T$, and up to 25000 cm$^{-1}$ at 300 K. In both cases, we took as reference the crystal itself, gold or silver-coated by \textit{in situ} evaporation, respectively. The results are shown in Fig.\ \ref{plane}-a. We assumed $R$(3.3 K) = 1 for $\omega \le 2\Delta_{ab}$ and we scaled\cite{Ortolani} the sub-THz $R_{ab}(T)$ at any $T$ according to the ratios in Fig.\ \ref{ratio-ab}.  Between 60 and 35 cm$^{-1}$  we interpolated the data with the Hagen-Rubens model. We thus obtained at any $\omega$ and $T$ the absolute reflectivity of the $ab$ planes, which was then used to extract the real part of the optical conductivity $\sigma_{ab}(\omega)$ by standard Kramers-Kronig transformations. At a first inspection of Fig.\ \ref{plane}-a, the plasma edge in the $R_{ab} (\omega)$ of CaAlSi falls at $\sim$ 8000 cm$^{-1}$, to be compared with\cite{Guritanu} $\sim$ 17000 cm$^{-1}$ in MgB$_2$.

The real part of the optical conductivity  $\sigma_{ab} (\omega)$ of the hexagonal planes is shown in Fig.\ \ref{plane}-b. For $T < T_c$ it decreases in the sub-THz range, showing the opening of the superconducting gap. As reported below, here $\Gamma_{ab} (T_c) \simeq$ 500 cm$^{-1} >> \Delta_{ab}(0)$ (dirty limit). Therefore, according to the FGT sum rule, the spectral weight $W_s$ condensed at $\omega$ = 0  at $T < T_c$ is\cite{DresselGruner}

\begin{equation}
W_s (T) = \int^{6\Delta}_{0+}[{\sigma}(10 K)-{\sigma} (T)]d\omega  \, .
\label{spectralweight}
\end{equation}

\noindent 
where $\sigma(T)$ is expressed in cm$^{-1}$ and it is assumed (see the inset of Fig.\ \ref{plane}-b) that $W_s$ saturates above $6\Delta_{ab} \simeq 60$ cm$^{-1}$.  This is therefore the range of validity of the FGT sum rule. Eq.\ \ref{spectralweight}
can be used to evaluate the field penetration depth $\lambda$. From $\lambda (T)= 1/(8W_s(T))^\frac{1}{2}$ one extracts $\lambda_{ab}$ (3.3 K)= 2250 $\pm$ 200 nm. As in the dirty limit \cite{DresselGruner}  $\lambda\sim\lambda^L(\Gamma/\Delta)^\frac{1}{2}$,  where $\lambda^L$ is the London penetration depth, one obtains $\lambda^L_{ab}$ (3.3 K) = 300 $\pm$ 30 nm, in excellent agreement with the $\lambda^L_{ab}$ (3.3 K) = 330 nm extracted from high-precision susceptibility measurements\cite{Prozorov}.
 
In the normal state, $\sigma_{ab} (\omega)$ shows a standard Drude behavior below about 1 eV, with a plasma frequency $\omega^0_{ab}$ = 20000 cm$^{-1}$ nearly independent of temperature.The same value comes out from Density Functional Theory calculations, after renormalization for the electron-phonon coupling\cite{private}. This value should be compared with $\omega^0_{ab} \simeq$ 45000 cm$^{-1}$ in\cite{Guritanu} MgB$_2$, and a scattering rate $\Gamma_{ab}$ = 480 cm$^{-1}$ at 10 K (850 cm$^{-1}$ at 300 K). 
Figure\ \ref{plane}-b also shows an unshielded phonon peak at 195 cm$^{-1}$, independent of $T$, which can be identified with the $E_{1u}$ in-plane stretching mode predicted between \cite{Mazin,Boeri,Huang} 183 and 194 cm$^{-1}$. The broad electronic band between 10000 and 20000 cm$^{-1}$ can be assigned to the transitions between the antibonding $\pi$ band and the $3d_{z^2-r^2}$ band of the Local Density Approximation (LDA) calculations\cite{Boeri}.

\begin{figure}[tbp]
    \epsfxsize=8cm \epsfbox {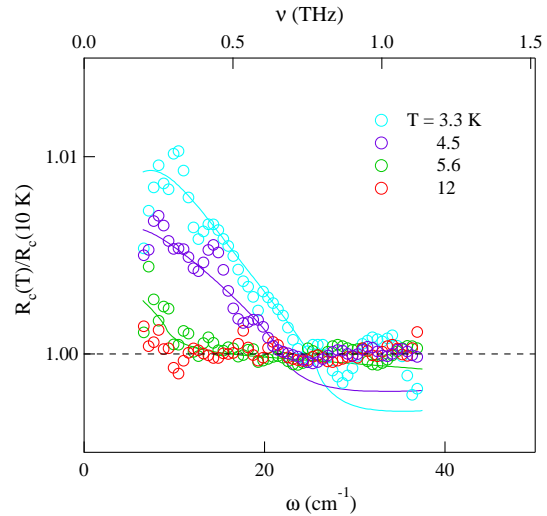}
     \caption{Color online. Ratio between the sub-THz reflectivity of the $c$ axis below $T_c$ and that in the normal phase at 10 K (circles). The lines are fits to a two-gap model, see Ref. \onlinecite{Ortolani3}.}
\label{ratio-c}
\end{figure}

\begin{figure}[tbp]
    \epsfxsize=8cm \epsfbox {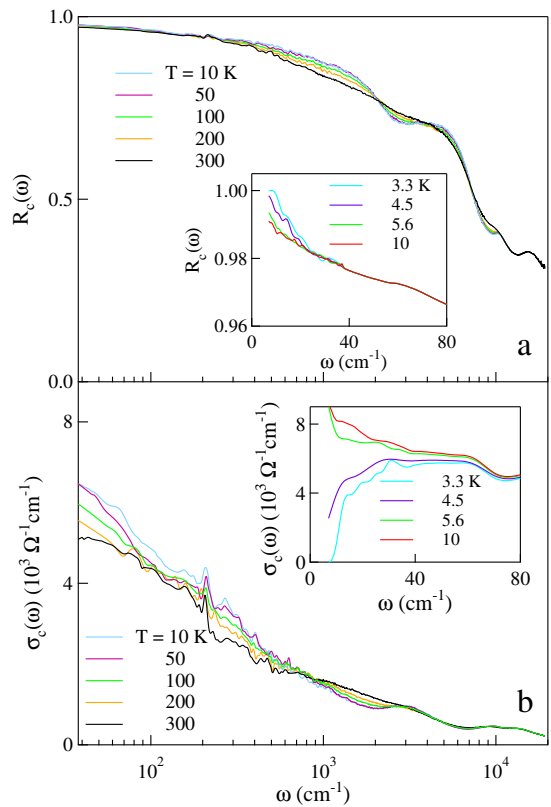}
       \caption{Color online. Optical response of the $c$ axis of CaAlSi: a) absolute reflectivity  at different temperatures. In the inset, reflectivity in the THz range above and below $T_c$, as reconstructed from the ratios in Fig.\ \ref{ratio-c}; b) optical conductivity obtained from the reflectivity in a. The inset shows the conductivity of the $c$ axis in the sub-THz range. }
\label{axis}
\end{figure}

Let us now consider the optical response of CaAlSi along the $c$ axis. Figure\ \ref{ratio-c} shows the ratio $R_{c,s}(T)/R_{c,n}$(10 K), as measured with the radiation polarized along the $c$ direction. Here also, the total variation of the reflectivity between 10 and 3.3 K is barely 1\%, due to the extremely high reflectivity in the normal phase (see Fig.\ \ref{axis}-a). A peak appears below $T_c$, as in Fig.\ \ref{ratio-ab}, but with a different shape. Such a shape indicates   that either only a  fraction of the carriers contribute to the optical conductivity of the SC phase, or  there are  two distinct gaps. The former case is observed for example in high-$T_c$ cuprates, where the order parameter has nodes in the $k$ space due to its $d$-wave symmetry, but this should be excluded in CaAlSi\cite{private}. The second situation is that of MgB$_2$, where a two-gap model well accounts for the reflectivity data below $T_c$\cite{Ortolani3}. In CaAlSi, the two gaps should come out from different regions of the Fermi surface, which is topologically disconnected along  the $k_z$ direction. By applying the same model\cite{Ortolani3} to the  $R_{c,s}(T)/R_{c,n}$(10 K) of Fig. \ \ref{ratio-c} one finds the fitting curves shown  in the Figure as solid lines. The resulting values for the two gaps are: $2hc\Delta_{c,1}$ = 22, 26, and 28  cm$^{-1}$ at 4.5, 3.3, and extrapolated to 0 K, respectively;
$2hc\Delta_{c,2}$ = 2, 6, and 8  cm$^{-1}$ at 4.5, 3.3, and 0 K, respectively. It is reasonable to associate the two gaps to the  bands which cross the Fermi level. According to the fit, the corresponding  plasma frequencies are  $\omega^0_{c,1}$ = 5500 cm$^{-1}$ and $\omega^0_{c,2}$ = 7100 cm$^{-1}$. Unlike in MgB$_2$, however, in CaAlSi both those bands have a  3D character and therefore one should expect two gaps also in the $ab$ plane . However, in that plane the Fermi surface is substantially connected topologically\cite{Boeri} and a one-gap model is expected to well fit the observations, as indeed shown in  Fig.\ \ref{ratio-ab}.
One may also remark that $\Delta _{ab} < \Delta_{c,1}$. Using here a higher energy resolution than in ARPES, we have  thus found  a moderate  anisotropy of the SC state which reflects the symmetry of the electron-phonon coupling \cite{Mazin,Boeri} and  may reconcile the ARPES observations with the penetration depth results of Ref. \onlinecite{Prozorov}. 

The absolute $R_c(T)$ was then determined for the $c$ axis as for the $ab$ plane, except for the need to interpolate data taken with different sources.  The resulting curves are shown in Fig. \ \ref{axis}-a, with the sub-THz range enlarged in the inset. The resulting conductivity $\sigma_c (\omega)$ is reported in Fig. \ \ref{axis}-b. At high energies, it exhibits shallow maxima at 4000, 9000, and 12000 cm$^{-1}$ which can be ascribed to the folding of the Brillouin zone resulting from the superstructures found by  X rays along the $c$ axis of our crystal.  The Drude term along the $c$ axis has an $\omega^0_{c}$ = 9000 cm$^{-1}$ (($\omega^0_{c})^2$ = ($\omega^0_{c,1})^2$ + ($\omega^0_{c,2})^2)$ and a $\Gamma_{c}$ of 180 cm$^{-1}$ at 10 K, 235 cm$^{-1}$ at 300 K. The phonon peak at 205 cm$^{-1}$ can be assigned to the $A_{2u}$ out-of-plane stretching mode, expected between\cite{Mazin,Boeri,Huang}  212 and 222 cm$^{-1}$. 
For $T < T_c$ $\sigma_{c} (\omega)$ decreases in the sub-THz range, showing the opening of the superconducting gap. As for the $ab$ plane, we estimated the spectral weight condensed below $T_c$ at $\omega$ = 0 by Eq.\ \ref{spectralweight}. This gives $\lambda_{c}$ (3.3 K) = 4000 $\pm$ 400 nm and  $\lambda^L_{c}$ (3.3 K) = 900 $\pm$ 100 nm. The value reported in Ref. \onlinecite{Prozorov}  for $\lambda^L_{c}$ at the same temperature is 725 nm.

In conclusion, we have studied the electrodynamics of the superconductor
CaAlSi from the sub-THz region to the visible range. The use of Coherent Synchrotron Radiation allowed us to measure the superconducting gap $2\Delta$ both in the hexagonal planes and along the $c$ axis. In both cases we find  $\Gamma (T_c) >> \Delta (0)$ and, as expected for a non-correlated system in the dirty limit, the FGT sum rule is fulfilled within $\sim 6\Delta$. In the planes we obtain good fits by a single gap value, while along $c$ one needs to introduce two gaps. This anisotropy is consistent with the calculated Fermi surface, which is connected in the $k_x, k_y$ planes, disconnected along the $k_z$ direction. All gaps values are consistent with BCS superconductivity in moderately strong coupling, but 2$\Delta_{ab}$ is slightly smaller than the largest 2$\Delta_{c}$.  This reflects also in the London penetration depth, with $\lambda^L_c \simeq 3 \lambda^L_{ab}$. This moderate anisotropy in the basic parameters of the SC state may partially explain, and possibly reconcile, the contrasting results reported in the literature of CaAlSi  by different techniques.
Finally, in the normal phase of CaAlSi, the Drude parameters $\omega_p$ and $\Gamma$ display an anisotropy not observed in MgB$_2$, being both much larger in the planes than along the $c$ axis. This suggests that the carriers propagating along $c$ have a larger effective mass, but longer free paths. A strong IR-active phonon is also observed, at a frequency very close to that calculated in the literature.

\end{document}